# [Bispropylurea Bridged Polysilsesquioxane: A Microporous MOF-like Material for Molecular Recognition](https://doi.org/10.1016/j.chemosphere.2021.130181)

Esmail Doustkhah,[a,*] Rafat Tahawy,[a] Ulla Simon,[b] Nao Tsunoji,[c] Yusuke Ide,[a] Dorian A. H. Hanaor,[b] M. Hussein N. Assadi[d,*]

[a] *International Center for Materials Nanoarchitechtonics (MANA), National Institute for Materials Science (NIMS), 1-1 Namiki, Tsukuba, Ibaraki 305-0044, Japan.*

[b] *Fachgebiet Keramische Werkstoffe, Technische Universität Berlin, 10623 Berlin, Germany.*

[c] *Graduate School of Advanced Science and Engineering, Applied Chemistry Program, Hiroshima University, 1-4-1 Kagamiyama, Higashi-Hiroshima 739-8527, Japan.*

[d] *School of Materials Science and Engineering, The University of New South Wales, Sydney, NSW, 2052, Australia.*

doustkhah.esmail@nims.go.jp, h.assadi.2008@ieee.org

## Abstract

Microporous organosilicas assembled from polysilsesquioxane (POSS) building blocks are promising materials that are yet to be explored in-depth. Here, we investigate the processing and molecular structure of bispropylurea bridged POSS (POSS-urea), synthesised through the acidic condensation of 1,3-bis(3-(triethoxysilyl)propyl)urea (BTPU). Experimentally, we show that POSS-urea has excellent functionality for molecular recognition toward acetonitrile with an adsorption level of 74 mmol/g, which compares favourably to MOFs and zeolites, with applications in volatile organic compounds (VOC). The acetonitrile adsorption capacity was 132-fold higher relative to adsorption capacity for toluene, which shows the pores are highly selective towards acetonitrile adsorption due to their size and arrangement. Theoretically, our tight-binding density functional and molecular dynamics calculations demonstrated that this BTPU based POSS is microporous with an irregular placement of the pores. Structural studies confirm maximal pore sizes of ~ 1 nm, with POSS cages possessing an approximate edge length of ~ 3.16 Å.

## 1. Introduction

The fabrication and implementation of nanoporous materials represent a highly active area of research towards applications, including adsorption (Zebardasti et al., 2020; Zhu et al., 2020), molecular recognition (Chen et al., 2010; Doustkhah and Ide, 2019), water treatment (Göksu et al., 2016; Yıldız et al., 2016; Arefi-Oskoui et al., 2019; Xu et al., 2019; Kobya et al., 2020), size-selective catalysis (Kataoka et al., 2015; Chen et al., 2018; Ide et al., 2018; Doustkhah et al., 2019; Akbari et al., 2020) and gas separation (Nugent et al., 2013; Kang et al., 2019).

Nanoporous materials include subsets of microporous and mesoporous materials; the term microporous refers specifically to materials exhibiting a typical pore size smaller than 2 nm, while mesoporous corresponds to pore sizes of 2–50 nm (Rouquerol et al., 1994). Microporous materials, in particular, tend to exhibit pores having diameters comparable in length-scale to the molecular size of solute compounds (Jeong et al., 2003; Ide et al., 2019). The most well-studied classes of such materials include metal-organic-frameworks (MOFs), zeolites, and engineered carbon (Ndamanisha and Guo, 2012; Zhang et al., 2017; Baumann et al., 2019; Doustkhah and Ide, 2020; Gang et al., 2021). Microporous materials with controlled structure are particularly of rapidly growing interest towards the applications listed above.

Among the broad landscape of microporous materials, polysilsesquioxanes (POSS) show significant unexplored functionality (Loy and Shea, 1995; Sato et al., 2018; Du and Liu, 2020; Zhang et al., 2020). In recent years a large number of studies into microporous POSS compounds have been reported. However, only a few structural analyses have been conducted as these materials do not usually crystallise in a manner similar to metal-organic-frameworks (MOFs), and thus, techniques such as X-ray diffraction and Raman spectroscopy are of limited use in their characterisation (Stein, 2003; Shimojima and Kuroda, 2020).

Despite difficulties in the synthesis and characterisation of POSS based microporous materials, this class of materials offers a chemically robust alternative to MOFs, which are generally unstable in acidic or basic conditions (Eddaoudi et al., 2000; Hu and Shea, 2011). POSS based microporous materials are highly stable under both acidic and basic conditions because the silsesquioxane cages are formed via covalent Si–O bonds, and are further connected to organic linkers via covalent Si–C bonds (Guan et al., 2019). Pores are created by the voids among the nanoscale organic bridges that link the POSS cages together (Ryu et al., 2020). In this regard, POSSs resemble MOFs, in which similar organic linkers link metal clusters instead of the silsesquioxane cages (Mofarah et al., 2019). Due to the flexibility of organic bridges in POSSs, pores in these materials are, however, not aligned—unlike the pores in MOFs. Since only a limited number of similarly structured silsesquioxane cages are used in POSSs, pores structures are mainly differentiated based on their bridging linkers. Here, we report the synthesis of a new urea-based POSS and examine its pore structure and performance in molecular recognition.

## 2. Methodology

For the material's synthesis, which is referred to as POSS-urea here, (3-aminopropyl)triethoxysilane (APTES) and (3-isocyanatopropyl)triethoxysilane (IPTES) were set to react under solvent-free conditions at 60 °C to yield BTPU. This organosiloxane precursor has been used in the past with promising results for the synthesis of hybrid networked mesoporous materials (Gao et al., 2009; Doustkhah et al., 2020). However, this work represents the first time POSS materials have been fabricated from this precursor. BTPU was then hydrolysed and condensed in an acidic (pH = 2) aqueous solution at 40 °C. Through this acidic condensation, the ethoxy groups at either end of the BTPU molecule bond together, resulting in a material comprising networked cage-like structures having Si vertices (Figure S1). Full experimental details are provided in the supplementary information.

Scanning electron microscopy (SEM) images were conducted using a Hitachi SU8000 field emission scanning electron microscope (FE-SEM). Transmission electron microscopy (TEM) and elemental mapping images were conducted using a JEM-2100F with an accelerating voltage of 200 kV. FT-IR spectra were obtained by a Shimadzu IR-460 spectrometer. Solid-state NMR spectra were recorded at 119.17 MHz on a Varian 600PS solid-state NMR spectrometer using a 6 mm diameter zirconia rotor. Surface area and pore size measurements, employing BET analysis of nitrogen sorption isotherms, were conducted by the Autosorb Belsorp-max instrument for POSS-urea. However, this technique was unable to identify the pore surface areas in this work due to the inability of nitrogen to access pore space at temperatures of 77 K.

Spin-polarised self-consistent tight-binding density functional (TBDF) calculations were performed using DFTB+ code (Hourahine et al., 2020) and the *pbc* element interactions library (Köhler et al., 2001), which produces accurate atomic geometries. To further improve the reliability of the calculations, the dispersion effects (Elstner et al., 2001) were explicitly included in the calculations. For the electronic setting, a fine Monkhorst-Pack *k*-point grid with a spacing of ~ 0.03 Å$^{-1}$ was used. Full geometry optimisation was carried out with convergence criteria for the energy and forces of 0.01 kcal/mol and 0.05 kcal/mol/Å, respectively. The building block of the POSS-urea obtained from TBDF calculations was then employed to construct a large nanoparticle for room temperature molecular dynamics simulation. Further details are provided in the supplementary information.

## 3. Results and discussion

### 3.1. Experimental characterisations

SEM and TEM images, given in Figure 1a and b, respectively, show typical spherical POSS-urea particles, with the latter exhibiting a diameter of 720 nm. The average diameter size of these spherical particles was ~ 1 μm (Figure 1c). The varying intensity inside the nanospheres, shown by the TEM image, indicates the ultralow density of the POSS-urea, measured at 0.11 g/cm$^3$, which is even lower than organically modified mesopores with a density of 0.655 g/cm$^3$ (Cao et al., 2020).

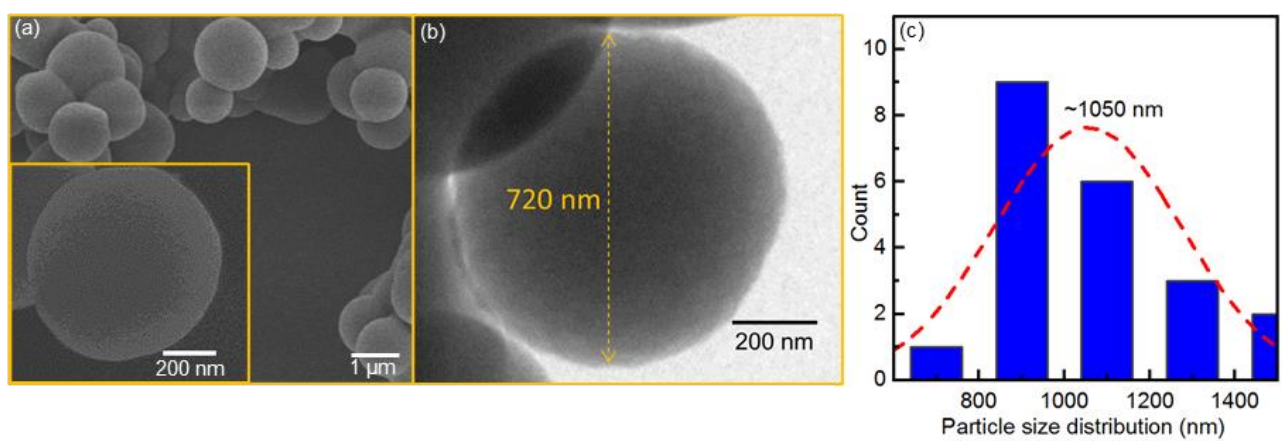

**Figure 1. (a) Low-magnified and high-magnified (inset) SEM and (b) TEM images of POSS-urea. (c) Particle size distribution calculated based on the SEM image. Although most particles had a diameter of ~ 800–900 nm, a few particles with diameters larger than 1 μm increased the average.**

As shown in Figure 2a, the solid-state magic angle spinning (MAS) $^{29}$Si-NMR spectrum of POSS-urea reveals only two species of silicon atoms, T$^3$ and T$^2$, with a ratio of 70:30, respectively. T$^3$ $^{29}$Si NMR peaks correspond to Si atoms having three Si–O bonds, which is characteristic of those Si atoms that constitute the vertices of POSS cages. The T$^3$/T$^2$ ratio indicates that at most 70% of silicon atoms are located in the structure of complete silsesquioxane cages (Figure S1a). The remaining ~ 30% of Si atoms of T$^2$ species can be ascribed to incomplete cages (Figure S1b), and to a lesser extent, to the surface silanol groups (Doustkhah et al., 2018; Protsak et al., 2019).

The $^1$H MAS NMR (Figure 2b) shows a peak at ~ 1.1 ppm that indicates the existence of the single silanols (Si–O–H), and once again, confirms the presence of incomplete cages (Figure S1b) (Protsak et al., 2019). The other two peaks at 5.1 ppm and 7.2 ppm can be attributed to the water-interacted silanols and urea hydrogens, respectively (Protsak et al., 2019).

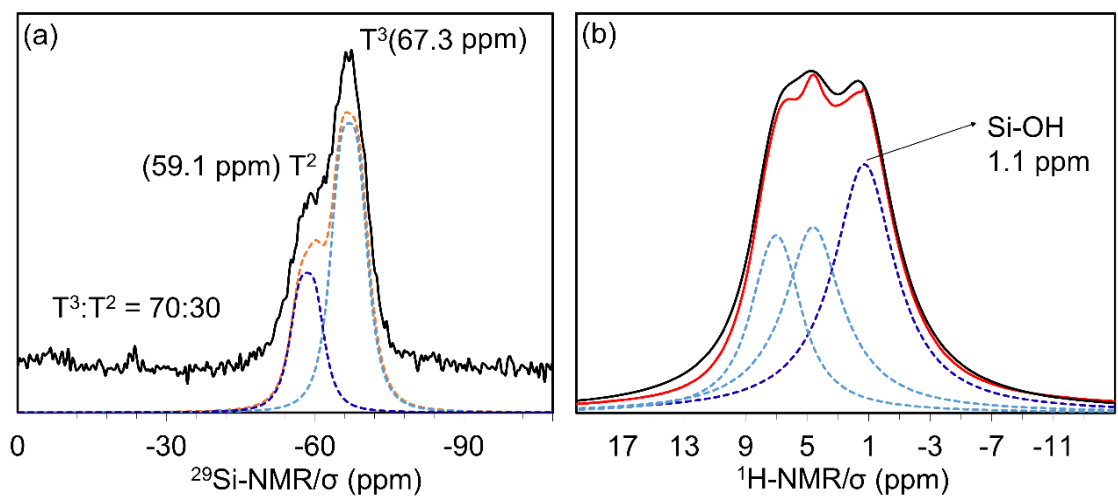

**Figure 2. (a) Solid-state $^{29}$Si MAS NMR and (b) Solid-state $^{1}$H MAS NMR spectra of POSS-urea.**

A comparison of the FTIR spectra of BTPU and POSS-urea (Figure 3a) shows that a new peak has appeared at 692 cm$^{-1}$ in the latter, which can be ascribed to the formation of T$^8$ silsesquioxane cages (Schäfer and Kickelbick, 2017). The two characteristic peaks at 1563 and 1627 cm$^{-1}$ further confirm that the BTPU's urea functionality has remained intact during the synthesis. The appearance of a peak at 1010 cm$^{-1}$ in POSS-urea also shows the Si–O–Si bond, which can confirm the condensation of the BTPU siloxanes. CHN analysis of the POSS-urea indicates that the obtained wt% ratio of the elements exhibits a good match with the expected results (details in the Supplementary Information). The $^{13}$C cross-polarisation (CP) MAS NMR (Figure 3b) also indicates that the urea carbonyl, as well as the whole BTPU chain, have remained intact during the hydrolysis and condensation steps (Bassindale et al., 2003).

Thermogravimetric analysis of POSS-urea shows 2.5 wt% loss in early 140 ºC corresponding to the trapped H$_2$O molecules in the material and 51.1 wt% loss due to the organic part decomposition and dehydration of silanols (Figure S2). The remaining material after thermal treatment up to 800 ºC can be attributed to the SiO$_2$ structure, which equals 46.4 wt%. These results are in good agreement with the wt% data of the POSS structure existing within POSS-urea. Therefore, the chemical composition of the POSS-urea can be presented as: Si$_{5.5}$C$_{19.3}$H$_{33.0}$O$_{15}$N$_{5.5}$.H$_2$O. The calculation based on the TGA shows that ~ 25% of Si in POSS-urea are in the form of silanol (Si–OH) of broken POSS cages, and the rest of Si can be attributed to the completely caged Si. This result is in agreement with the $^{29}$Si MAS NMR findings (Figure 2a).

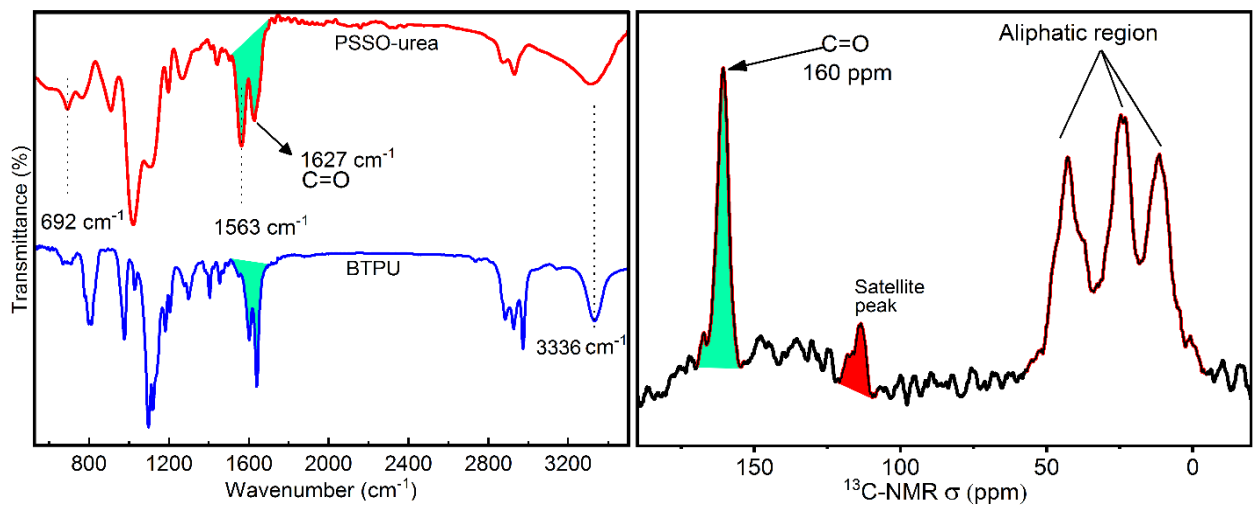

**Figure 3. (a) FTIR spectra of BTPU and POSS-urea and (b) $^{13}$C CP MAS NMR of POSS-urea.**

The applicability of the BET method for measuring the micropores' surface area of the pure POSSs (Figure S3), which has no secondary pores or nano-space and possess a short-length bridging organic moiety (*e.g.*, bispropylurea) is problematic (Centeno and Stoeckli, 2010; Hu et al., 2020). The problem arises from the fact that BET can sometimes indicate either larger or smaller pore size and surface area than actual values for the microporous materials depending on many factors (Gelb and Gubbins, 1998; Thommes et al., 2015; Gómez-Gualdrón et al., 2016; Tian and Wu, 2018; Datar et al., 2020). The discrepancy between the BET estimation and the actual geometric area is more common and more severe when the pore sizes approach ~ 2 nm. The underestimation mainly originates because of the limited space inside the pores, which forces $N_2$ to condense before forming a uniform monolayer, an assumption on which the BET measurement is based on (Bardestani et al., 2019; Sinha et al., 2019). This condition particularly applies to the POSS-urea synthesised here as demonstrated by molecular dynamics calculations at 77 K (Figure S4).

Doustkhah, Esmail, et al. 2021, "Bispropylurea bridged polysilsesquioxane: A microporous MOF-like material for molecular recognition." *Chemosphere,* 276, 130181.

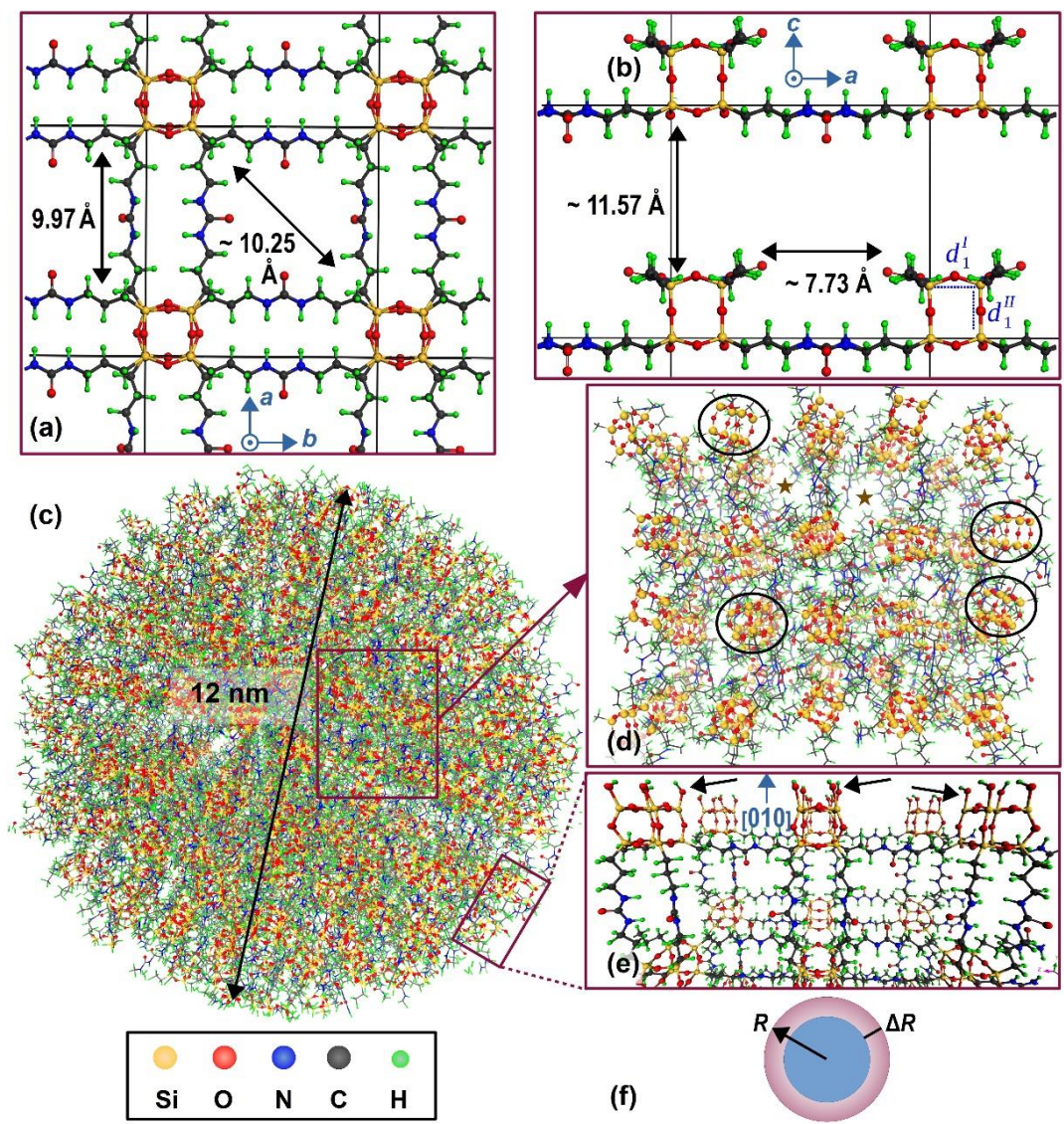

**Figure 4. (a) Top view, (b) side view of -urea the building block of POSS obtained from tight-binding density functional calculations. (c) Spherical nanoparticle built from the POSS building block and equilibrated with molecular dynamics simulation. (d) Magnified snapshot of the bulk area of the POSS-urea nanoparticle. (e) Magnified snapshot of the atomic layers at the surface of the POSS-urea nanoparticle. (f) Schematics for calculating the surface to volume ratio.**

### 3.2. Computational Study

To gain insight into the structure and pore geometry of POSS-urea materials, we combined simulations based on TBDF calculations with molecular dynamics analysis pertaining to ambient conditions. In the TBDF simulations, we probed the POSS-urea structure that exclusively contains fully formed $T_8$ cages (Figure S1a), which is the simplest structure formed by POSS-urea. This choice is justified by the fact that $T_8$ cages are the most stable, and thus the most prevalent in the POSS structure (Hossain et al., 2008; Rehman and Gwaltney, 2014). The absence of any other silicon-containing precursor other than BTPU dictates that there must be one bispropylurea bridge for every two Si atoms in the studied materials. As a result, the basic and most stable polymeric unit manifests as (4bispropylurea-$Si_8O_{12}$)$_n$. Here, Si atoms form the cube-like $Si_8O_{12}$ POSS cage with one Si atom at each corner, which are connected by oxygens along the edges of the cages—this is the geometry of $T_8$ cage. Furthermore, every Si atom is at one end of a bispropylurea bridge connecting it to another Si atom in an adjacent POSS cage. As a result, there are 3 Si–O and 1 Si–C bonds at every vertex of the POSS cage, resulting in the $T^3$ Si atoms detected by the $^{29}$Si NMR. In our case, because of the synthesis protocol that preserves the initial Si–C bonds, every bridge links two corners (Si atoms) from two neighbouring POSS cages directly, without the occurrence of doubly-chained bridges, thus forming a space-filling network of bridged and slightly distorted cubes. In addition to the stoichiometric restrictions, the POSS cages linked by these bispropylurea bridges should also create a repeatable unit to fill the space for successful solid formation when fully polymerised. One possible such polymerisation network, which we examined for stability by TBDF calculations, is shown in Figure 4. Here, the POSS-urea forms a 2D network in which bispropylurea bridges link the POSS cubes within a basal plane (Figure 4a), with no cross planes linkers, as having cross-plane linkers would not conform to the stoichiometric restriction. The polymerised planes are then stacked to form a 3D solid (Figure 4b). This structure in POSCAR format and its converted crystallographic information file (CIF) are supplied in the supplementary information of this article. Because the bispropylurea bridges are flexible, we do not expect that the POSS-urea planes remain flat at ambient conditions.

Because the TBDF calculations pertain to 0 K conditions, Figure 4a,b present a structure in which the length of the bispropylurea bridge is fully stretched, as bending the bridges requires thermal energy input (Fakirov, 2017). In the fully stretched structure, the edges, *i.e.* the length of the Si–O–Si chain, of the POSS cage are calculated as 3.16 Å along the bispropylurea bridges ($d_1^I$ in Figure 4b), and 3.31 Å along the stacking direction ($d_1^{II}$ in Figure 4b). Furthermore, the length of the fully stretched bridge between Si atoms is 9.97 Å, and the interlayer distance is 11.57 Å. This arrangement creates voids with diameters of ~ 10 Å in the space between the POSS cages that can be utilised as molecular sieves for adsorption and recognition applications. However, at room temperature, the bridges are expected to bend and twist, so as these voids may contract and distort.

To study the room-temperature nanoscale behaviour of the POSS-urea, we constructed a sizeable spherical nanoparticle of 12 nm diameter from the building blocks of Figure 4a,b for molecular dynamics simulation (the largest size possible for tractable simulation on our machine). The equilibrated structure is shown in Figure 4c, while Figure 4d presents a snapshot of the interior of the nanoparticles in which the POSS cages are visible (marked with black circles). Additionally, a considerable percentage of the voids were also conserved in the equilibrated structure (marked with stars in Figure 4d). The preservation of the POSS cages was confirmed by the analysis of the radial distribution function (RDF). The Si RDF (Figure 5) of the equilibrated nanoparticle shows three peaks with descending order at 3.05, 4.31 and 5.27 Å. The first peak ($d_1$) corresponds to the nearest neighbour Si atoms of Si–O–Si chains along the cage edges. The second peak at $d_2$ corresponds to the second nearest neighbour Si atoms along the POSS cage side diagonal, while the third peak at $d_3$ corresponds to the third nearest neighbour Si atoms along the POSS body diagonal. For higher *r* values, there was no sharp RDF peak but rather a wide and shallow shoulder (marked with an arrow). The lack of any Si RDF peaks at higher distances indicates significant bending and twisting of the bispropylurea bridges that have disrupted any long-range order among the POSS cages.

Within the interior of the simulated POSS-urea nanoparticle, all Si atoms are trifunctional (T variety) as all the POSS cage corners are linked by the bispropylurea bridges and hence have one Si–C bond. On the surface of the nanoparticle, however, tetrafunctional Si (Q variety) may occur due to surface cleavage. As in Figure 4e, the outermost atomic layer of the POSS-urea nanoparticle contains POSS cages that are capped with OH group instead of being linked to a bispropylurea bridge. Consequently, the surface to volume ratio of the POSS-urea nanoparticle directly determines the ratio of the Q to T Si atoms. Therefore, we can estimate the Q to T ratio of the Si atoms by calculating the fraction of the volume of a thin shell containing the outmost atomic layers to the volume of the remainder of the nanoparticle (Figure 4f), assuming a homogenous distribution of the POSS cages within the nanoparticle and on the surface. For a spherical nanoparticle, the shell volume is $4\pi R^2 \Delta R$. Here, *R* is the radius of the nanoparticle and $\Delta R$ is the thickness of the outermost shell. The remainder of the volume of the nanoparticle is $\frac{4}{3}\pi R^3 - 4\pi R^2 \Delta R$. Consequently, the ratio is Q to T Si atoms is:

$$\frac{Q}{T} = \frac{1}{2} \cdot \frac{R^2 \Delta R}{1/3 R^3 - R^2 \Delta R}. \qquad (1)$$

The ½ factor is applied as only half of the Si atoms at the surface shell of the nanoparticle are of Q variety (marked with black arrows in Figure 4e). Assuming $\Delta R$ is 0.3 nm—the side length of the POSS cage. For a nanoparticle of ~ 700 nm, the Q to T ratio is ~ 0.1% which is quite negligible. Expectedly, no Q peaks were visible in the $^{29}$Si spectrum of Figure 2a. This statement can also be generalised for the $T^2$ Si atoms that may form on the surface.

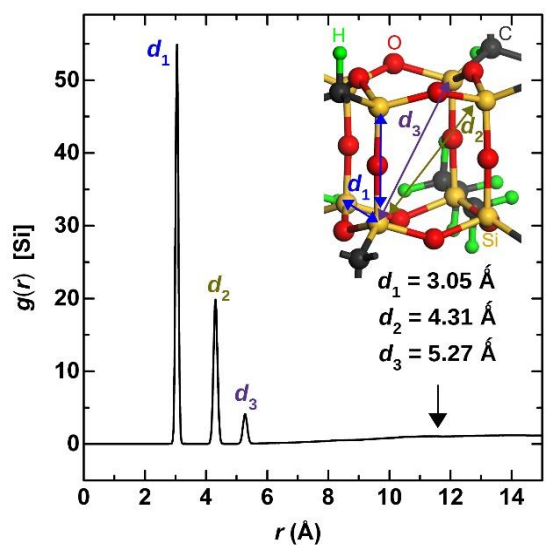

**Figure 5. Si radial distribution function of the equilibrated POSS-urea nanoparticle.**

### 3.3. Material Performance

To examine material performance, we evaluated the pores' behaviour in molecular recognition tests. Accordingly, we selected different compounds with small molecular size with different functional groups and sizes, *i.e.*, acetonitrile, methanol, and toluene. We treated POSS-urea with acetonitrile, toluene, and methanol at a concentration of 2000 ppm. At an adsorption capacity of 74 mmol/g at room temperature, acetonitrile, with a smaller and linear molecular structure (at a length of ~ 3.28 Å), was most efficiently adsorbed into POSS-urea (Figure 6). The observed adsorption capacity of POSS-urea towards acetonitrile is significantly larger than values found for synthetic clays and zeolites, which are reported to adsorb up to ca. 10 mmol/g (Florian and Kubelkova, 1994; Urita et al., 2019).

Surprisingly, methanol, which is also considered a small molecule, having a length of ~ 2.4 Å, did not show any detectable adsorption into the POSS-urea. The lack of methanol adsorption may be attributed to intermolecular hydrogen bonding—based on proton exchange—between methanol molecules, which results in bulkier combinations that diffuse only weakly into the pores (Borho et al., 2006). This behaviour contrasts with that of acetonitrile that generally interacts via neutral hydrogen bonds with zeolites' surface (Florian and Kubelkova, 1994). Toluene, the largest molecule examined (with a diameter up to 6.8 Å), with no possible intermolecular hydrogen bonding, was nonetheless more diffusible than methanol in adsorption. Relative to acetonitrile, however, the adsorption capacity of toluene at 0.58 mmol/g was ~132-fold lower (Figure 6a). Toluene's lack of adsorption is due to the larger volumetric single molecular size (6.8 Å) relative to acetonitrile (Qie et al., 2019). Additionally, the adsorption capacity of acetonitrile increased with rising temperature (Figure 6a). Based on the Arrhenius plot and equation (Figure S6), the activation energy for acetonitrile was 41 kJ/mol. The rate constants ($k$) were calculated by the pseudo-first-order model, according to the following equation (Doustkhah and Ide, 2019; Henning et al., 2019):

$$\text{Ln}(q_{\max} - q_t) = \text{Ln} q_{\max} - kt$$
(2)

where $q_{max}$ indicates the maximum adsorption capacity (mg/g), $q_t$ represents the quantity of adsorbed molecules (mg/g) at time $t$, and $k$ shows rate constant of pseudo-first-order adsorption (g/mg·min).

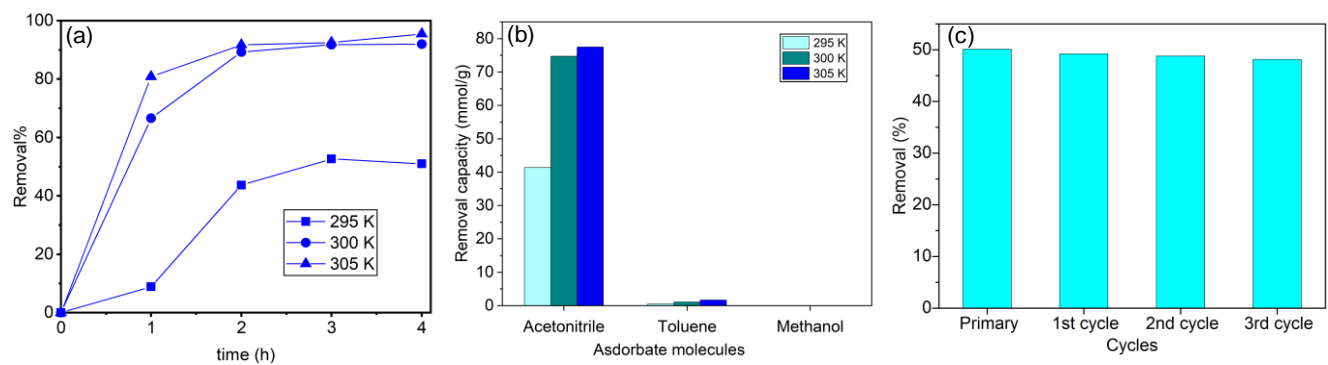

**Figure 6.** (a) Adsorption capacity POSS-urea toward acetonitrile, toluene, and methanol at 295 K, 300 K, and 305 K. (b) Time course adsorption of acetonitrile (2000 ppm) on POSS-urea at different temperatures at pH = 9. (c) Recyclability of POSS-urea at room temperature for three consecutive cycles.

## 4. Conclusions

In conclusion, we have demonstrated the synthesis of a novel microporous material, *i.e.*, POSS-urea, with the chemical composition of $Si_{5.5}C_{19.3}H_{33.0}O_{15}N_{5.5}\cdot H_2O$ and average particle size of ~ 1µm, and characterised its structure across multiple length scales. POSS-urea formed here exhibit high molecular recognition capability among three tested organic molecules ($CH_3CN$, MeOH, and toluene) that vary in size and functional groups. Through theoretical calculations, we have elucidated the structure of the POSS cages, and their bispropylurea bridged assembly in three dimensions. The results of this work demonstrate the potential capability of POSSs of functional and stable microporous materials and motivate further investigations into organosiloxane based systems.

## Conflicts of interest

There are no conflicts of interest to declare.

## Acknowledgements

The authors gratefully acknowledge the funding of this project by computing time provided by the Paderborn Center for Parallel Computing (PC²). Further computational resources were provided by the Integrated Materials Design Centre at The University of New South Wales. We would like to thank Sören Selve and Jan R. J. Simke for TEM imaging (ZELMI, TU Berlin).

# Supplementary Information

## 1.     Chemicals

(3-aminopropyl)triethoxysilane (APTES) 98% and (3-isocyanatopropyl)triethoxysilane (IPTES) 95% were purchased from Wako Co., Japan.

## 2. Materials Synthesis

Synthesis of 1,3-bis(3-(triethoxysilyl)propyl)urea (BTPU) was performed according to the literature.[1] In brief, the APTES and IPTES with the molar ratio of 1:1 were set to react under solvent-free conditions at 60 °C for 1 h. In the end, the obtained solid was the final product. For the synthesis of POSS-urea (**Figure S1**), the obtained BTPU (12 mmol, 7.75 g) was dissolved in an aqueous solution containing triblock copolymer (2g, P123) and stirred vigorously. The pH of the solution was arranged at 2 with HCl, and the temperature was set at 40 °C. After 2 h, the reaction mixture was transferred to a Teflon-lined stainless-steel autoclave and aged at 100 °C for 24 h. Next, the mixture was filtered, and the solid was washed for 5 times with EtOH (50 mL) and three times with chloroform (20 mL), and finally, dried at 60 °C for 12 h. CHN analysis of the final product was obtained as follow: C:27.4; H:4.86; N:9.2. TG-DT analyses of POSS-urea was recorded using a Rigaku Thermo plus TG8120 apparatus. For measuring the density, 10 mg of POSS-urea was placed in a 5 ml balloon. The balloon was then fully filled by adding toluene. The density was calculated by subtracting the volume of the added toluene and obtaining the POSS-urea volume in the balloon.

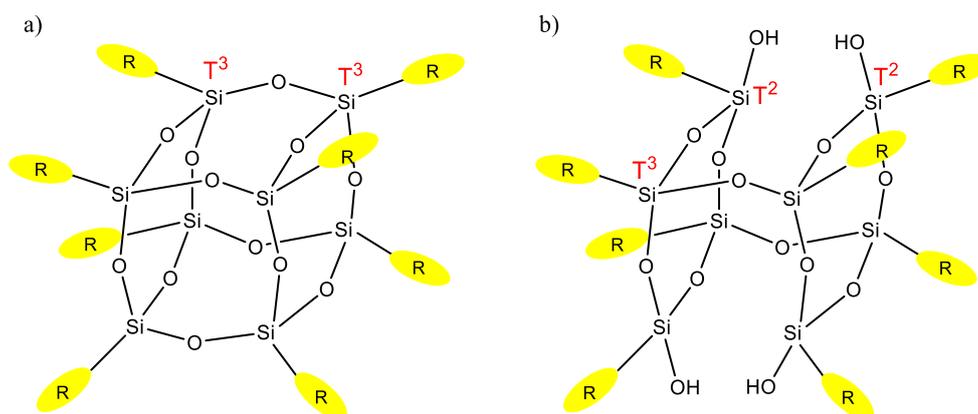

**Figure S1**. (a) and (b) show the schematic representation of a complete and a broken octa(urea-based) polysilsesquioxane cages, respectively. $T^3$ and $T^2$ siloxanes are marked accordingly. In the broken cages, both $T^2$ and $T^3$ varieties exist while in complete cages, only $T^3$ siloxanes can exist.

Doustkhah, Esmail, et al. 2021, "Bispropylurea bridged polysilsesquioxane: A microporous MOF-like material for molecular recognition." *Chemosphere,* 276, 130181.

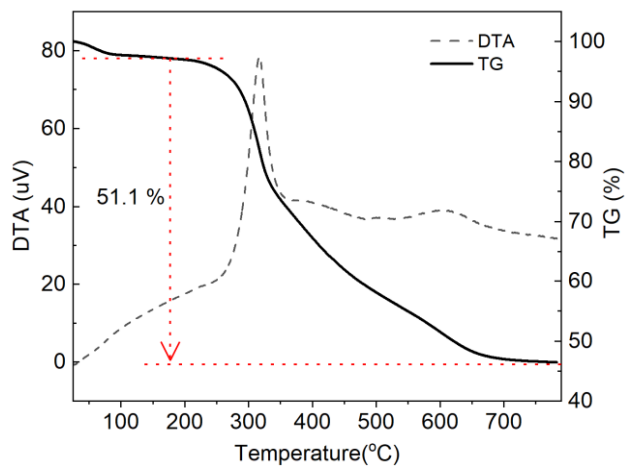

**Figure S2**. TG-DT analysis of POSS-urea in $O_2$ atmosphere.

### 3.   BET Characterisation

The BET characterisation, shown in **Figure S3**, did not detect any microporosity despite the evidence of microporosity from acetonitrile adsorption capacity. As seen in **Figures 4a** and **4b** of the article, the maximum theoretical pore diameter is ~ 1 nm. In the equilibrated structure at room temperature (**Figure 4c**), due to the flexibility and relatively short length of the urea bridge which is shown in **Figure S4a**, the pore diameter can be slightly smaller at ~ 0.9 nm. Molecular dynamics studies conducted at 77 K confirmed the thermal contraction of pores to about half of that size at places within the POSS-urea nanoparticle which consequently results in the closure of the pores to nitrogen adsorption as illustrated in **Figure S4b**. Furthermore, the existence of broken cages and entrapped solvents intensify the difficulty of obtaining reliable BET results. The inability of nitrogen to access pores in such microporous materials at 77K has been observed elsewhere and is the subject of several discussions in the literature.[2-4]

Doustkhah, Esmail, et al. 2021, "Bispropylurea bridged polysilsesquioxane: A microporous MOF-like material for molecular recognition." *Chemosphere,* 276, 130181.

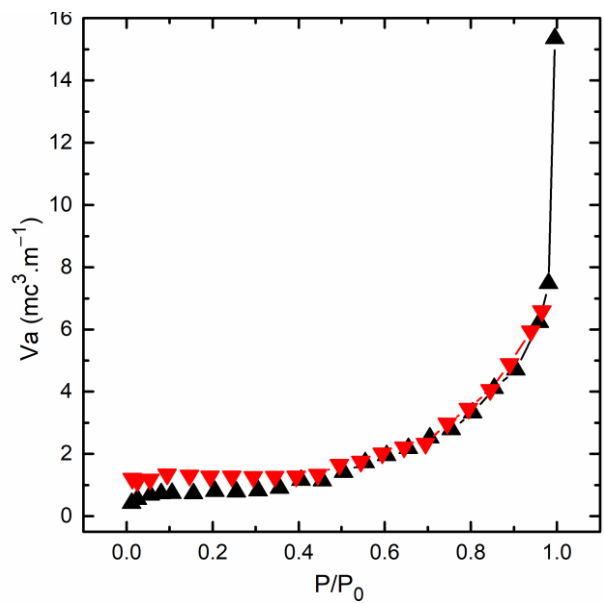

**Figure S3.** N$_2$ adsorption-desorption profile for POSS-urea at 77 K.

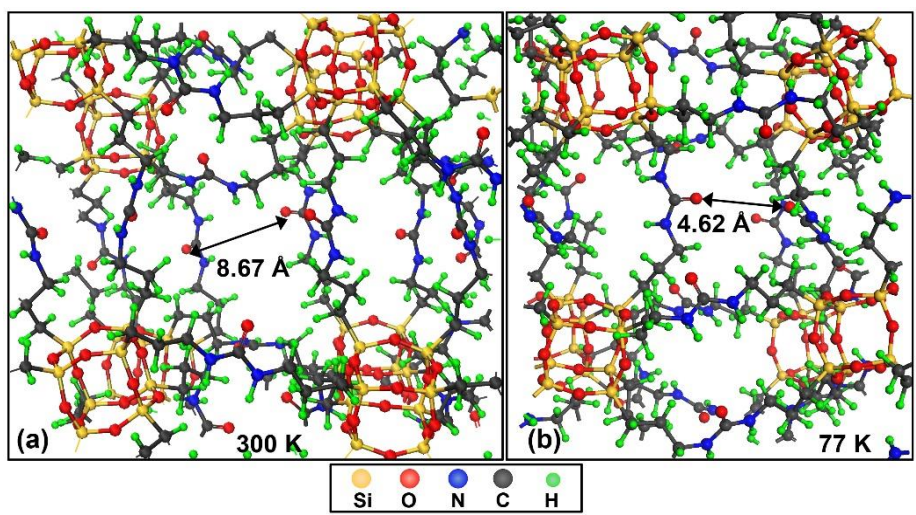

**Figure S4.** Two snapshots of the molecular dynamics run at two temperatures for the POSS-urea nanoparticle (initially shown in **Figure 4c**). The left panel pertains to simulation at 300 K, while the right panel to 77 K, the condition of the BET experiment. The snapshots were taken from precisely the same atoms within the nanoparticle. These snapshots indicate that the shrinkage in pore size can be severe in the BET experiment condition, rendering the BET result unapplicable to POSS-urea.

## 4. Tight-Binding Density Functional Calculations

For simulating the structure of POSS-urea, a T$_8$ POSS cage, constructed according to the previously calculated structure,[5,6] was placed at a corner of three-dimensional supercell and connected by the urea bridges from corner to corner, conforming to the stoichiometry of the synthesised POSS-urea. The structure was then fully optimised both in terms of atomic coordinates and lattice parameters, using the quasi-Newton method. A few possible configurations were considered of which the most stable

Doustkhah, Esmail, et al. 2021, "Bispropylurea bridged polysilsesquioxane: A microporous MOF-like material for molecular recognition." *Chemosphere,* 276, 130181.

one was presented. The construction and validation methods of the urea connected organosilica structures have been explained in more details in the literature.[1] Finally, to examine the stability of the predicted structure, an *ab initio* molecular dynamics run based on the density functional tight-binding method was performed on a supercell of the $2a \times 2b \times 2c$ of the optimised structure using NPT ensemble with Nose thermostat[7] and Anderson barostat[8] at $T$ = 300 K for 5 ps with steps of 1 fs.[9,10] All bonds were conserved, indicating stability at ambient condition.

### 5. Molecular Dynamics Calculations

Based on the POSS-urea unit cell, of which the CIF is also included in the SI, optimised by tight-binding density functional calculations, a large spherical nanoparticle of a diameter of 12 nm was constructed. This nanoparticle contained 64562 atoms for which the dangling bonds at the surface were capped with H. The nanoparticle was then enclosed with a large cubic box of 20 nm edges. Then a molecular dynamics run were performed using the highly-scalable LAMMPS package[11] and the readily available COMPASS force field parameter data.[12,13] The COMPASS force field has been successfully and widely applied for studying the organosilica systems.[14-16] The large box containing the POSS-urea nanoparticle was allowed to equilibrate under NVT (constant mass, volume, and temperature) ensemble at 300 K for 0.5 ns with steps of 0.5 fs. The simulation was once more run at 77 K pertaining to the conditions of the BET experiment. Although the volume of the simulation box was fixed, the nanoparticle itself slightly shrank during the run resulting in a higher density for the nanoparticle relative to the unequilibrated structure. The values for temperature and energy were monitored to ensure they had minimum fluctuations. At the final steps of the run, the fluctuations were ~ 0.5% of the total values. The simulated XRD of the equilibrated POSS-urea nanoparticle at 300 K is shown in **Figure S5**. The first peak at 1.26° is generated by the regular nanoparticle arrangement imposed by the periodic lattice boundary. The second peak at 1.78° indicates that although the pores are not aligned perfectly, they possess a certain degree of mesoporosity.[17,18] The lack of any significant diffraction peaks at higher angles indicates the POSS-urea, forming the pore walls, is amorphous.

Doustkhah, Esmail, et al. 2021, "Bispropylurea bridged polysilsesquioxane: A microporous MOF-like material for molecular recognition." *Chemosphere,* 276, 130181.

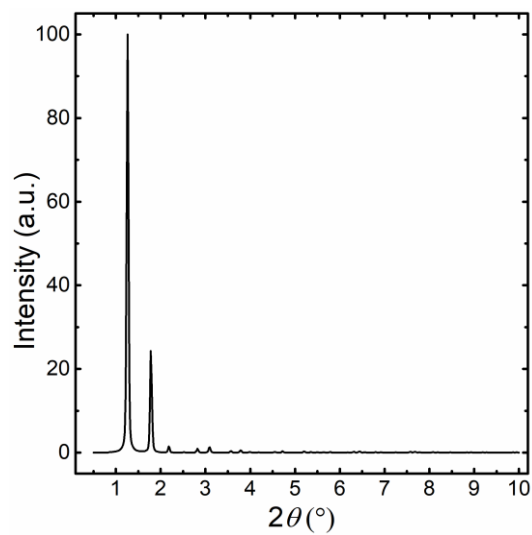

**Figure S5.** Simulated low angle XRD of the equilibrated POSS-urea nanoparticle.

## 6. Molecular Recognition Test

The sample powder (15 mg) as an adsorbent was added to a 20 ml solution from a mixture of $H_2O$ (85 mL, pH = 9) and 15 ml of methanol containing acetonitrile (2000 ppm) in a septum-sealed glass vial and sonicated for 30 seconds and stirred on a temperature-controlled magnetic stirrer at room temperature for three hours. In the end, the sample powder was filtered by a syringe filter of 0.22 μm membrane. The filtrate was analysed by a gas chromatograph (Shimadzu BID-2010 Plus GC), and the adsorbed amount of acetonitrile and methanol on the adsorbent was quantified accordingly. In the case of toluene adsorption, experimental detail is the same, except toluene was used instead of the aqueous solution.

## 7. Adsorption Tests

In this section, the Arrhenius plot for acetonitrile adsorption based on Equation 2 of the manuscript is presented. The activation energy was calculated to be 41 kJ/mol.

Doustkhah, Esmail, et al. 2021, "Bispropylurea bridged polysilsesquioxane: A microporous MOF-like material for molecular recognition." *Chemosphere,* 276, 130181.

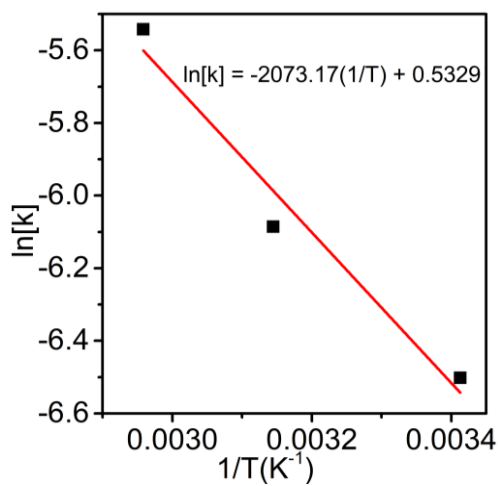

**Figure S6.** Arrhenius plot for adsorption of acetonitrile on POSS-urea through three different temperatures, *i.e.*, 295 K, 300 K, 305 K.

## 8. The Optimized Structure of POSS-urea

Below is the structure of the POSS-urea as optimized by our tight-binding density functional calculations in POSCAR format. We have also included a conversion of this structure into CIF. CIF, although not the most suitable format for calculated structures, is the most widely used among experimentalist and theoreticians alike, and as such, can be visualized by many software packages and can be easily converted to other formats.

```
POSS-urea
1.0
     14.4113998413        0.0000000000        0.0000000000
      0.0000000000       15.9294004440        0.0000000000
      0.0000000000        0.0000000000       15.9294004440
   O    Si    C    N    H
  16     8   28    8   56
Direct
     0.882080019         0.196109995         0.197770000
     0.886669993         0.002310000         0.192059994
     0.883679986         0.197889999         0.996159971
     0.888260007         0.000400000         0.001670000
     0.030750036         0.100720003         0.231099993
     0.031880021         0.100720003         0.965309978
     0.731519997         0.095459998         0.200560004
     0.734369993         0.095519997         0.991970003
     0.033139944         0.205369994         0.098210000
     0.737709999         0.228640005         0.096380003
     0.041550040         0.996460021         0.098229997
```

Doustkhah, Esmail, et al. 2021, "Bispropylurea bridged polysilsesquioxane: A microporous MOF-like material for molecular recognition." *Chemosphere,* 276, 130181.

```
0.740760028        0.962540030        0.096330002
0.056489944        0.600300014        0.357890010
0.715709984        0.835839987        0.596480012
0.143380046        0.599900007        0.899119973
0.689289987        0.351570010        0.596539974
0.996900022        0.197339997        0.197779998
0.772069991        0.997009993        0.193250000
0.768899977        0.195800006        0.998989999
0.002959967        0.002970000        0.998979986
0.998499990        0.197650000        0.998120010
0.773580015        0.996739984        0.999710023
0.001309991        0.003300000        0.196899995
0.767340004        0.195490003        0.194049999
0.041210055        0.600430012        0.273059994
0.041980028        0.287499994        0.259799987
0.013249993        0.372729987        0.224040002
0.053429961        0.446220011        0.273710012
0.045480013        0.913420022        0.259790003
0.016199946        0.828360021        0.224010006
0.055209994        0.754530013        0.273939997
0.734480023        0.920099974        0.596440017
0.729359984        0.933860004        0.283540010
0.760900021        0.968479991        0.368420005
0.720790029        0.919529974        0.442400008
0.730769992        0.933759987        0.909330010
0.761590004        0.968689978        0.824370027
0.721580029        0.919480026        0.750519991
0.054859996        0.600380003        0.930339992
0.046049953        0.287680000        0.937359989
0.010059953        0.372819990        0.967599988
0.052070022        0.446159989        0.918879986
0.049720049        0.913230002        0.937330008
0.013010025        0.828320026        0.967519999
0.053900003        0.754630029        0.918569982
0.718829989        0.269870013        0.596490026
0.717570007        0.255730003        0.283399999
0.752250016        0.224570006        0.368640006
0.705510020        0.268920004        0.442339987
0.719099998        0.255919993        0.909560025
0.753149986        0.224439994        0.824240029
0.706459999        0.269100010        0.750689983
0.030779958        0.674669981        0.233480006
0.029739976        0.526329994        0.233390003
0.747179985        0.959280014        0.670610011
0.746890008        0.959270000        0.522210002
```

Doustkhah, Esmail, et al. 2021, "Bispropylurea bridged polysilsesquioxane: A microporous MOF-like material for molecular recognition." *Chemosphere,* 276, 130181.

```
0.014780045        0.674799979        0.948189974
0.013900042        0.526409984        0.948440015
0.736559987        0.232290000        0.670639992
0.736169994        0.232240006        0.522279978
0.030850053        0.673579991        0.168200001
0.029979944        0.527530015        0.168109998
0.017369986        0.282009989        0.325159997
0.118370056        0.283730000        0.262479991
0.936720014        0.377669990        0.222859994
0.036579967        0.377710015        0.158030003
0.130270004        0.439599991        0.278219998
0.026149988        0.446660012        0.338730007
0.020380020        0.919309974        0.324970007
0.121880054        0.916710019        0.262910008
0.039960027        0.823099971        0.158140004
0.939620018        0.824069977        0.222360000
0.027500033        0.754390001        0.338820010
0.132079959        0.760389984        0.278849989
0.749589980        0.024550000        0.669570029
0.749279976        0.024540000        0.523259997
0.652869999        0.931730032        0.281560004
0.753300011        0.868440032        0.276639998
0.739619970        0.035000000        0.374119997
0.837539971        0.967819989        0.372000009
0.746420026        0.853950024        0.442369998
0.643760026        0.916610003        0.436890006
0.654299974        0.931209981        0.911650002
0.755159974        0.868440032        0.915970027
0.838220000        0.968599975        0.820469975
0.739690006        0.035050001        0.818809986
0.644580007        0.916159987        0.756229997
0.747619987        0.854040027        0.750549972
0.944239974        0.674049973        0.961939991
0.943340003        0.527890027        0.962050021
0.122460008        0.286579996        0.941560030
0.028980017        0.279549986        0.870320022
0.025589943        0.380670011        0.035170000
0.933610022        0.374960005        0.961619973
0.037909985        0.438360006        0.850539982
0.128419995        0.447580010        0.926670015
0.126150012        0.913770020        0.941229999
0.032320023        0.921779990        0.870419979
0.936489999        0.826900005        0.961839974
0.028749943        0.820219994        0.035020001
0.130259991        0.752380013        0.926129997
```

Doustkhah, Esmail, et al. 2021, "Bispropylurea bridged polysilsesquioxane: A microporous MOF-like material for molecular recognition." *Chemosphere,* 276, 130181.

```
0.039579988        0.762650013        0.850279987
0.747489989        0.167740002        0.669570029
0.747089982        0.167689994        0.523370028
0.734300017        0.322919995        0.276149988
0.641219974        0.251029998        0.281109989
0.828279972        0.233080000        0.372880012
0.739419997        0.156289995        0.374190003
0.628679991        0.263509989        0.436419994
0.722320020        0.336789995        0.442409992
0.736270010        0.323040009        0.916589975
0.642729998        0.251610011        0.912159979
0.739690006        0.156250000        0.818780005
0.829219997        0.232400000        0.819700003
0.723770022        0.336870015        0.750620008
0.629610002        0.264160007        0.756839991
```

---

Doustkhah, Esmail, et al. 2021, "Bispropylurea bridged polysilsesquioxane: A microporous MOF-like material for molecular recognition." *Chemosphere,* 276, 130181.

S18    Rao, K. M., Parambadath, S., Kumar, A., Ha, C.-S. & Han, S. S. Tunable Intracellular Degradable Periodic Mesoporous Organosilica Hybrid Nanoparticles for Doxorubicin Drug Delivery in Cancer Cells. *ACS Biomater. Sci. Eng.* **4**, 175–183 (2018). https://doi.org/10.1021/acsbiomaterials.7b00558

25